\documentclass[
a4paper,															% Benutzt DIN-A4-Papier-Format (=default)
11pt,		 														% Schriftgröße
parskip=on,  														% seitlicher Abzug neuer Absatz on/off
oneside, 															% twoside = zweiseitiger Satz
onecolumn,															% Zweispaltig (typisches Paper-Format)
numbers=noenddot,													% Die Abschnittsnummerierungen werden immer ohne den letzten Punkt dargestellt, also "1.2.3" statt "1.2.3."
bibliography=totoc,													% Bindet das Literaturverzeichnis ins Inhaltsverzeichnis ein
listof=totoc,														% Bindet das Tabellenverzeichnis und Abbildungsverzeichnis ins Inhaltsverzeichnis ein (falls vorhanden)
]{scrartcl} 														% Dokumentenklasse Article

\usepackage[german, english]{babel} 			% Überschriften werden von LaTeX in der korrekten Sprache erzeugt																				

\usepackage[utf8]{inputenc}						% Ermöglicht Eingabe von Umlauten 																					
																				
\usepackage{babelbib}							% um Bibtex mit deutschen Styles zu verwenden

\usepackage[T1]{fontenc}		        		% Schriftart	
											
\usepackage{lmodern}							% Zur Verwendung von Sonderzeichen (z.B allgemeine Gaskonstante)%%%früher:dsfont

\usepackage{amsmath} 					  		% Paket mit vielen Formelumgebungen, u.a. align

\usepackage{amssymb} 							% für die Menge der natürlichen, rationalen,... Zahlen. Befehl z.B. \mathbb{N}

\usepackage[version=3]{mhchem}					% Fü chemische Formeln und chemische Gleichungen. Bespiel zur Verwendung: \ce{CH4 + 2 O2 <-> CO2 + 2 H2O}

\usepackage{float}								%Paket für Gleitumgebung und erzwingt Positionierung von Floatobjekten mit "`H"'

\usepackage{graphicx}			          		% Ermöglicht die Einbindung von Bildern

\usepackage[figuresright]{rotating}				% Damit können Bilder, Tabellen, usw. in landscape Format angezeigt werden

\usepackage{xspace} 							% Abstände nach selbst definierten Befehlen können richtig gestaltet werden

\usepackage[dvipsnames]{xcolor}					% Zur Verwendung von Farben

\usepackage[numbers, sort&compress]{natbib}		% modify citation style

\usepackage{multicol} 						% Ermöglicht flexiblen Einsatz mehrerer Spalten

\usepackage[verbose]{placeins} 				% Stellt den Befehl \floatbarrier zur Verfügung. Dieser erzwingt die 
\usepackage[babel]{microtype} 				% das Paket wird an die beim Paket babel getroffene Spracheinstellung angepasst
\usepackage[								% Seitenränder						
left=28mm,  %28								% Breite des linken Rands
right=25mm,	%25								% Breite des rechten Rands
bottom=25mm,								% Breite des unteren Rands
top=28mm									% Breite des oberen Rands
]{geometry} 

\usepackage{fancyhdr}
\usepackage{cite}

\usepackage[labelfont=bf]{caption} 	% für fette Labels in float-Objekten

\usepackage{setspace}
\usepackage{boldline}	 			% for thick table grid lines

\usepackage{siunitx}
\usepackage{lscape} 		% für Tabellen im Querformat
\usepackage{pdflscape}		% damit Seiten in landscape-Modus auch als landscape im PDF-reader angezeigt werden
\usepackage{tablefootnote} 	% für Fußnoten in Tabellen. 
\usepackage{multirow} 		% mehrere Zeilen in Tabellen zusammenfassen
\usepackage{longtable}	
\usepackage{arydshln} 		% für gepunktete Linien in Tabellen
\newcolumntype{C}[1]{>{\centering\arraybackslash}p{#1}}
\usepackage{booktabs}		% für \toprule und \bottomrule in Tabellen
\usepackage{bm} 			% Fette Buchstaben im Mathe-Modus mit \bm
\usepackage{mathtools} 		% für := (Definition) mittels \coloneqq
\usepackage{supertabular}
\usepackage{tabularx} 		% ermöglicht Tabellen mit automatischen Zeilenumbrüchen, wenn ein Zeileneintrag zu lang wird. Dafür eine Spalte mit X (statt l, r oder c) erstellen
\usepackage[para]{threeparttable}
\usepackage{etoolbox} 
\AtEndEnvironment{threeparttable}{\vskip10pt}{}{}% change here

\allowdisplaybreaks 	% ermöglicht Seitenumbrüche bei langen align-Umgebungen

\usepackage{bookmark}	% für Lesezeichen im PDF

\write18{} 	% Für shell-escape
\usepackage{tikz}
\usepackage{pgfplots} 	% Lädt auch TikZ
\pgfplotsset{compat=newest} % läd neueste Version
\usepgfplotslibrary{external,polar,patchplots}%fillbetween}
\tikzset{external/force remake} % Forciert die neukompilierung aller Abbildungen, wenn zuvor bereits ein PDF mit demselben Namen (über \tikzsetnextfilename) erzeugt wurde

\usetikzlibrary{
	arrows,
	patterns,
	calc,
	shapes.arrows,
	intersections,
	decorations,
	shapes,
	positioning,
	shadings,
	plotmarks,
	pgfplots.colormaps, 
	spy 		% erlaubt das Erstellen von Lupen-Ansichten.
} % Benutzen vorgefertigter TikZ Ressurcen
\newcommand{\cotwo}{CO\textsubscript{2}\xspace}		% CO2
\newcommand{\chfour}{CH\textsubscript{4}\xspace}	% CH4
\def\code#1{\texttt{#1}}    % enables direct formating into typical code font
\usepackage{makecell} 		% für Zeilenumbrüche in Tabellen-Zellen

\usepackage{hyphenat}
\begin{document}
																												
%%%EINBINDEN DES INHALTS
%\pagestyle{plain} 	% keine headings, aber Seitenzahlen

% alle Voreinstellungen von Headern & Footern zurücksetzen:
\pagestyle{fancyplain} 	% keine Seiten-Nummerierung
\fancyhf{} 	
\renewcommand{\headrulewidth}{0pt}	% keine Linie zur Abtrennung von Seiteninhalt und Headern/Footern
\chead{ \fancyplain{}{This work has been accepted for the 22nd European Control Conference.
} } 	% zentraler Header 

\begin{center}
\huge Comparison of Unscented Kalman Filter Design for Agricultural Anaerobic Digestion Model \\ [5mm]

\Large
Simon Hellmann\textsuperscript{1,2}, Terrance Wilms\textsuperscript{3}, Stefan Streif\textsuperscript{2}, Sören Weinrich\textsuperscript{4,1} \\ [3mm]

\large
\textsuperscript{1} DBFZ Deutsches Biomasseforschungszentrum gGmbH, Leipzig, Germany,\\ simon.hellmann@dbfz.de\\ [3mm]

\textsuperscript{2} Chemnitz University of Technology, Chemnitz, Germany,\\ 
stefan.streif@etit.tu-chemnitz.de\\ [3mm]

\textsuperscript{3} Technische Universität Berlin, Berlin, Germany, \\ 
terrance.wilms@tu-berlin.de \\ [3mm]

\textsuperscript{4} Münster University of Applied Sciences, Münster, Germany, \\ weinrich@fh-muenster.de
\end{center}

\Large
This is the extended version of a paper published in the proceedings of the 22nd European Control Conference, held in Stockholm on June 25-28, 2024. Since July 27, 2024, it can be found under the following DOI: 10.23919/ECC64448.2024.10591126. The extended version is available under a CC BY-NC-ND 4.0 license.

\vfill
\normalsize

\copyright 2024 IEEE. Personal use of this material is permitted. Permission from IEEE must be obtained for all other uses, in any current or future media, including reprinting/republishing this material for advertising or promotional purposes, creating new collective works, for resale or redistribution to servers or lists, or reuse of any copyrighted component of this work in other works.

\newpage
\pagestyle{plain} 		% Seitennummerierung erlaubt, aber weder header noch footer
\setcounter{page}{1}
\cfoot[]{\thepage}		% Seitennummerierung 

\section*{Abstract}
\addcontentsline{toc}{section}{Abstract}
Dynamic operation of biological processes, such as anaerobic digestion (AD), requires reliable process monitoring to guarantee stable operating conditions at all times. Unscented Kalman filters (UKF) are an established tool for nonlinear state estimation, and there exist numerous variants of UKF implementations, treating state constraints, improvements of numerical performance and different noise cases. So far, however, a unified comparison of proposed methods emphasizing the algorithmic details is lacking. The present study thus examines multiple unconstrained and constrained UKF variants, addresses aspects crucial for direct implementation and applies them to a simplified AD model. The constrained UKF considering additive noise delivered the most accurate state estimations. The long run time of the underlying optimization could be vastly reduced through pre-calculated gradients and Hessian of the associated cost function, as well as by reformulation of the cost function as a quadratic program. However, unconstrained UKF variants showed lower run times at competitive estimation accuracy. This study provides useful advice to practitioners working with nonlinear Kalman filters by paying close attention to algorithmic details and modifications crucial for successful implementation. 

\begin{tabular}{rl}
	\textbf{Key words:} & Process monitoring, nonlinear state estimation, sigma point Kalman filter,\\ & biogas technology, ADM1
\end{tabular}

\clearpage

\section{Introduction}
Anaerobic digestion (AD) is an established technology for the treatment of biogenic waste. In the AD process, organic matter is converted into biogas \citep{Theuerl2019}. Demand-driven operation of biological processes such as AD requires reliable process monitoring to ensure stable operation \citep{Kazemi2020}. As a means of process monitoring, Kalman filters have been examined in numerous studies for model-based online state estimation in various domains \citep{Gentsch.2020, Cantelobre.2020}. In particular, the Unscented Kalman Filter (UKF) could be shown to be well suited for state estimation of nonlinear biological processes \citep{Raeyatdoost.2023, Kemmer.2023, Tuveri.2021}. 

To this end, Kolas et al. (2009) \citep{Kolas.2009} investigated various implementations of the UKF, involving different noise scenarios (additive and non-additive) as well as state constraints. More specifically, state constraints were addressed by adopting a nonlinear program (NLP) proposed by \citep{Vachhani.2006}, and by reformulating the NLP as a quadratic program (QP) assuming linear output equations. 

Weinrich and Nelles (2021) have recently proposed simplified AD models \citep{Weinrich2021b} derived from the highly complex Anaerobic Digestion Model No. 1 (ADM1) \citep{Batstone2002}. The potential of these ADM1 simplifications has been demonstrated in case studies addressing demand-oriented biogas production in lab- \citep{Weinrich2021} and full-scale \citep{Mauky.2016}. Moreover, the ADM1 simplifications have been shown to be locally observable, and thus appropriate to be applied in state estimation \citep{Hellmann.2023b}.

This paper compares different UKF designs for a simplified ADM1 model which is derived from \citep{Weinrich2021b}. By demonstrating multiple UKF implementations, we aim to provide insights into comparative performance of available algorithms, and thereby offer analytical, numerical and algorithmic guidance. The study thus also contributes to realizing model-based monitoring and control for demand-driven operation of AD plants.

\section{Unscented Kalman Filtering}
In this work, we consider discrete-time stochastic systems %of ordinary difference equations 
% of the form  
% 
% \begin{subequations}
% \label{eq:meth:systemEquationsFullyAug}
% \begin{IEEEeqnarray}{ll}
%     x_{k+1} &= f\left(x_k,u_k,w_k\right), \quad x_0 - \text{given} \label{eq:meth:systemEquationsFullyAug_f}\\
%     % 
%     y_k &= h\left(x_k,v_k\right) % \IEEEyesnumber
% \end{IEEEeqnarray}
% \end{subequations}
%
\begin{subequations}
\label{eq:meth:systemEquationsAdd}
%\begin{IEEEeqnarray}{ll}
\begin{align}
    x_{k+1} &= f\left(x_k,u_k\right) + v_k, \quad x_0 - \text{given} \label{eq:meth:systemEquationsAdd_f}\\
    y_k &= h\left(x_k\right) + w_k.
    \label{eq:meth:systemEquationsAdd_h}
\end{align}
%\end{IEEEeqnarray}
\end{subequations}
with state variables $x \in \mathbb{R}^{n}$,  control variables $u \in \mathbb{R}^{p}$, and measurement variables $y \in \mathbb{R}^{q}$. %$k$ represents the discrete time instant $t_k = k \Delta t$ with regular sample time $\Delta t$.
$f$ can also represent the integration of continuous-time differential equations on a discrete time grid $t_k = k\, \Delta t$ with sample time $\Delta t$ and $k \in \mathbb{N}_0$, see \citep{Vachhani.2006}. Process and measurement noise ($v \in \mathbb{R}^{n}$ and $w \in \mathbb{R}^{q}$) are assumed Gaussian and zero-mean with 
\begin{subequations}
%\begin{IEEEeqnarray}{rl}
\begin{align}
E\{v(k)\} &= 0, E\{w(k)\} = 0, \quad \forall{k}\\
E\{v(k) v^T(l)\} &= Q(k) \delta_{k,l} , \\
E\{w(k) w^T(l)\} &= R(k) \delta_{k,l} .
\end{align}
%\end{IEEEeqnarray}
\end{subequations}
$Q$ and $R$ are process and measurement noise covariance matrices and $\delta_{k,l}$ is the Kronecker delta. \eqref{eq:meth:systemEquationsAdd} shows the additive noise case. In case of non-additive noise, $v_k$ and $w_k$ are direct arguments of $f$ and $h$, i.e., $f\left(x_k,u_k,v_k\right)$ and $h\left(x_k,w_k\right)$. 
%Process and measurement noise $v_k$ and $w_k$ are shown in additive form in \eqref{eq:meth:systemEquationsAdd}. In case they are assumed non-additive, $v_k$ and $w_k$ act as additional independent variables of $f$ and $h$, respectively.
%  
The nominal linear time-variant equivalent of the nonlinear output equation \eqref{eq:meth:systemEquationsAdd_h} is denoted as
\begin{equation}
\label{eq:meth:linearOutput}
y_k = C_k x_k .
\end{equation}
% 
%
%In Kalman filtering, estimates of the state $\hat x_k$ are obtained and recursively updated both through the system model and noisy measurements 
%resulting in a two-step procedure consisting of time and measurement update. 
% To this end, 
Sigma point Kalman filters such as the UKF use scaled copies of the old estimate $\hat x_{k-1}$ called sigma points to predict a-priori estimates $\hat x_k^-$ (time update). These are in turn corrected with measurements $y_k$ to deliver a-posteriori estimates $\hat x_k$ (measurement update). 
%Analogously, the old state error covariance matrix $P_{k-1}$ undergoes time update delivering the prior $P_k^-$, and measurement update to yield the posterior $P_k$. 
The analogous procedure is applied for calculation of the state error covariance matrix $P_{k-1}$ with corresponding prior $P_k^-$ and posterior $P_k$. Each time step involves sigma points to be sampled from a scaled multivariate normal distribution with mean $\hat x_{k-1}$ and covariance matrix proportionate to $P_{k-1}$ \citep{vanderMerwe.2004b}.
%; then to be propagated in time and aggregated. 
%
\subsection{Unconstrained Case}
The basic concepts of the conventional unconstrained UKF are briefly summarized here. Sigma points $\chi_i$ are sampled around the state estimate $\hat x$ % and saved as
% % 
% \begin{equation}
% \label{eq:meth:sigmapointsSampling}
% \chi_{k-1} = \begin{bmatrix} \hat x_{k-1}, & \hat x_{k-1} + \gamma \sqrt{P_{k-1}}, & \hat x_{k-1} - \gamma \sqrt{P_{k-1}} \end{bmatrix}.
% \end{equation}
% % 
% {\color{red} [XY for Terrance/Sören: soll ich das hier einmal zeigen, wie die Sigmapunkte überhaupt gesampled werden, oder weglassen?]} 
which involves a scaling factor $\gamma$ \citep{vanderMerwe.2004b}, given by
\begin{subequations}
\label{eq:meth:UKF-scaling}
%\begin{IEEEeqnarray}{ll}
\begin{align}
\gamma &= \sqrt{n+\lambda} \quad \text{with}\\
\lambda &= \alpha^2 (n + \kappa) - n.
\end{align}
%\end{IEEEeqnarray}
\end{subequations}
For Gaussian noise, nominal tuning parameter values are recommended by \citep{Kolas.2009} as
\begin{equation}
\label{eq:meth:defaultTuning}
\begin{bmatrix} \alpha & \beta & \kappa\end{bmatrix} = \begin{bmatrix} 1 & 2 & 0 \end{bmatrix}.
\end{equation}
During time and measurement update, sigma points are aggregated as a weighted average with weights $W^x$ for states and $W^c$ for the state error covariance matrix \citep{vanderMerwe.2004b}
\begin{subequations}
\label{eq:meth:UKF-weights}
%\begin{IEEEeqnarray}{ll}
\begin{align}
W_0^x &= \lambda/(n+\lambda), \\
W_0^c &= \lambda/(n+\lambda) + 1 - \alpha^2 + \beta, \\
W_i^x &= W_i^c = 1/\left(2(n + \lambda)\right), \quad i = 1\ldots 2n.
\end{align}
%\end{IEEEeqnarray}
\end{subequations}
\paragraph{Augmentation}
Non-additive process noise is incorporated by extending the matrix $P$ with the process noise covariance matrix $Q$. Thereby the system state is augmented with zero-mean process noise \citep{Kolas.2009} (denoted with superscript index $a$). This results in an augmented system order $L=2n$
% 
% \begin{subequations}
% \label{eq:meth:augmentation}
% \begin{align}
% P^a_k &= \begin{bmatrix}
%     P_k & 0 \\
%     0 & Q_k \end{bmatrix} \\
% % 
% x^a_k &= \begin{bmatrix} x_k^T & 0 \end{bmatrix}^T \\
% %
% \chi_k &= \begin{bmatrix} (\chi_k^x)^T & (\chi_k^w)^T \end{bmatrix}^T
% \end{align}
% \end{subequations}
% 
\begin{subequations}
\label{eq:meth:augmentation}
% \begin{IEEEeqnarray}{ll}
\begin{align}
P^a_{k-1} &= \begin{bmatrix}
    P_{k-1} & 0 \\
    0 & Q_{k-1} \end{bmatrix}, \\ 
x^a_{k-1} &= \begin{bmatrix} x_{k-1} \\ 0 \end{bmatrix}, \,
\chi_{k-1} = \begin{bmatrix} \chi_{k-1}^x \\ \chi_{k-1}^v \end{bmatrix}.
\end{align}
% \end{IEEEeqnarray}
\end{subequations}
Analogously, non-additive zero-mean process and measurement noise is incorporated by extending $P$ with the process and measurement noise covariance matrices $Q$ and $R$. This results in the fully augmented system order $L=2n+q$
% 
% \begin{subequations}
% \label{eq:meth:fulAugmentation}
% \begin{align}
% P^a_k &= \begin{bmatrix}
%     P_k & 0 & 0\\
%     0 & Q_k & 0\\
%     0 & 0 & R_k\end{bmatrix} \\
% % 
% x^a_k &= \begin{bmatrix} x_k^T & 0 & 0 \end{bmatrix}^T \\
% % 
% \chi_k &= \begin{bmatrix} (\chi_k^x)^T & (\chi_k^w)^T & (\chi_k^v)^T \end{bmatrix}^T
% \end{align}
% \end{subequations}
%
\begin{subequations}
\label{eq:meth:fulAugmentation}
% \begin{IEEEeqnarray}{ll}
\begin{align}
\hspace{-6pt} P^a_{k-1} &= \begin{bmatrix}
    P_{k-1} & 0 & 0\\
    0 & Q_{k-1} & 0\\
    0 & 0 & R_{k-1}\end{bmatrix}, \hspace{1pt} 
    x^a_{k-1} = \left[x_{k-1}^T \hspace{3pt} 0 \hspace{3pt} 0 \right]^T, \\
%
% x^a_{k-1} &= \begin{bmatrix} x_{k-1} & 0 & 0 \end{bmatrix}^T,% \hspace{5pt}
% 
% \chi_{k-1} = \begin{bmatrix} \chi_{k-1}^x & \chi_{k-1}^v & \chi_{k-1}^w \end{bmatrix}^T.
\hspace{-6pt} \chi_{k-1} &= \left[(\chi_{k-1}^x)^T \hspace{5pt} (\chi_{k-1}^v)^T \hspace{5pt} (\chi_{k-1}^w)^T \right]^T.
\end{align}
% \end{IEEEeqnarray}
\end{subequations}
\normalsize
The distinction between nominal and augmented system order $n$ and $L$ is not addressed in \citep{Kolas.2009} and is therefore clarified here. % see also \citep{vanderMerwe.2004b} 
For the augmented and fully augmented version, computation of the weights \eqref{eq:meth:UKF-weights} must be adjusted for the increased system order $L$. This ensures that the aggregation from sigma points to estimates is maintained properly. By contrast, computation of the scaling factor \eqref{eq:meth:UKF-scaling} must still be conducted with the nominal system order $n$. Otherwise the effect of $P_k$ during sampling of sigma points deviates from the additive noise version. 

In \citep{Kolas.2009} a numerically more robust reformulation is proposed for computing the a-posteriori estimates $\hat x_k$ and $P_k$. This involves a separate update of the sigmapoint priors  $\chi_k^{x-}$ through the innovation
\begin{equation}
\label{eq:meth:sigmapointUpdateReformulated}
\chi_{k,i}^x = \chi_{k,i}^{x-} + K_k \left( y_k - h(\chi_{k,i}^{x-} , \chi_{k,i}^w) \right), \quad i=0\ldots 2L
\end{equation}
and then to aggregate them to posteriors $\hat x_k$ and $P_k$ 
\begin{subequations}
\label{eq:meth:measurementUpdateReformulated_x_P}
%\begin{IEEEeqnarray}{ll}
\begin{align}
\hat x_k &= \sum_{i=0}^{2L} W_i^x \chi_{k,i}^x \\
P_k &= \sum_{i=0}^{2L} W_i^c \left( \chi_{k,i}^x - \hat x_k \right)\left( \chi_{k,i}^x - \hat x_k \right)^T.
\end{align}
%\end{IEEEeqnarray}
\end{subequations}
The derivation was conducted for the fully augmented case in \citep{Kolas.2009}. For additive and augmented noise cases, $h(\chi_{k,i}^{x-} , \chi_{k,i}^w)$ in \eqref{eq:meth:sigmapointUpdateReformulated} must be replaced with $h(\chi_{k,i}^{x-})$. Further, computation of $P_k$ must be slightly adjusted for additive noise
% 
% \begin{align}
% \begin{split}
% P_{k+1} &= \sum_{i=0}^{2L} W_i^c \left( \chi_{k,i}^x - \hat x_k \right)\left( \chi_{k,i}^x - \hat x_k \right)^T + \\ 
% & \hspace{1em} Q_k + K_k R_k K_k^T
% \end{split}
% \end{align}
%
%\begin{displaymath}
\begin{equation}
P_k = \sum_{i=0}^{2L} W_i^c \left( \chi_{k,i}^x - \hat x_k \right)\left( \chi_{k,i}^x - \hat x_k \right)^T + Q_k + K_k R_k K_k^T
\end{equation}
%\end{displaymath}
% 
\normalsize
with $K_k$ according to \citep[Tab. 5]{Kolas.2009}. For augmented process noise, $P_k$ is correctly computed as
%
%\begin{displaymath}
\begin{equation}
%\small
P_k = \sum_{i=0}^{2L} W_i^c \left( \chi_{k,i}^x - \hat x_k \right)\left( \chi_{k,i}^x - \hat x_k \right)^T + K_k R_k K_k^T
\end{equation}
%\end{displaymath}
% 
\normalsize
with $K_k$ according to \citep[Tab. 6]{Kolas.2009}. 
\subsection{Constrained Case}
Upper and lower bounds on a-posteriori estimates can be accounted for by projection methods \citep{Simon.2005} and clipping, which can be done in various locations throughout each iteration of the UKF \citep{Kolas.2009}. To account for nonlinear inequality constraints %on a-posteriori estimates, 
\citep{Kolas.2009} adopted the NLP proposed by \citep{Vachhani.2006} but suggested to leave the scaling factor and weights as in \eqref{eq:meth:UKF-scaling} and \eqref{eq:meth:UKF-weights}, resulting in 
\begin{subequations}
%\begin{IEEEeqnarray}{ll}
\begin{align}
\label{eq:meth:optimizationProblem}
    \chi_{k,i}^x &= \arg \min_{\chi_{k,i}^x} J_{k,i}^{NLP} \\ 
    \text{s.t. } %x_{i,\text{low}} \leq \chi_{k,i}^x &\leq x_{i,\text{up}}, \quad 
    % A_i \chi_{k,i}^x &\leq b_i, \quad \text{where}
    &\tilde C (\chi_{k,i}^x) \leq 0, \quad \text{where} \label{eq:meth:optimizationProblemInequalities}
\end{align}
%\end{IEEEeqnarray}
\end{subequations}
\begin{equation}
\begin{split}
\label{eq:meth:costFunctionNLP}
J_{k,i}^{NLP} &= \left(y_k - h(\chi_{k,i}^x)\right)^T R_k^{-1} \left(y_k - h(\chi_{k,i}^x)\right) + \\
&\hspace{12pt} \left(\chi_{k,i}^x - \chi_{k,i}^{x-}\right)^T \left(P_k^-\right)^{-1} \left(\chi_{k,i}^x - \chi_{k,i}^{x-}\right). 
\end{split}
\end{equation}
The scope of this study is limited to linear inequality constraints. Therefore the nonlinear inequalities in \eqref{eq:meth:optimizationProblemInequalities} read
\begin{equation}
\label{eq:meth:linearInequalities}
\tilde C(\chi_{k,i}^x)  = A \chi_{k,i}^x - b \leq 0
\end{equation}
with $\tilde C$ and $b \in \mathbb{R}^{m \times 1}$, $A \in \mathbb{R}^{m \times n}$ and $m$ as the number of constraints. Upper and lower bounds on a-posteriori estimates and sigma points can be included in \eqref{eq:meth:linearInequalities}. 
For linear output equations \eqref{eq:meth:linearOutput}, \citep{Kolas.2009} further showed that the NLP cost function \eqref{eq:meth:costFunctionNLP} can be recast into the QP-form
\begin{equation}
\begin{split}
\label{eq:meth:costFunctionQP}
J_{k,i}^{QP} &= (\chi_{k,i}^x)^T \left( C_k^T R_k^{-1} C_k + \left( P_k^-\right)^{-1} \right) \chi_{k,i}^x \, + \\
&\hspace{12pt} - 2 \left( y_k^T R_k^{-1} C_k + ( \chi_{k,i}^{x-} )^T \left( P_k^-\right)^{-1} \right) \chi_{k,i}^x.
\end{split}
\end{equation}
The NLP was solved in Matlab by using \code{fmincon}, and by using \code{quadprog} for the QP.

\textbf{Remark:} Note that \eqref{eq:meth:costFunctionNLP} was proposed by \citep{Vachhani.2006} for the additive noise case, but \citep{Kolas.2009} adopted it for the augmented noise cases also. Since \eqref{eq:meth:costFunctionQP} was derived in \citep{Kolas.2009} considering the nominal output equation $h(x_k)$ for additive measurement noise, \eqref{eq:meth:costFunctionQP} also holds for additive noise. Lastly, as stated in \citep{Vachhani.2006} the posteriors $\hat x_k$ and $P_k$ must be computed from the solution of the NLP/QP as per \eqref{eq:meth:measurementUpdateReformulated_x_P}, since the measurement noise covariance matrix $R$ is already considered in the QP/NLP, see \eqref{eq:meth:costFunctionNLP} and \eqref{eq:meth:costFunctionQP}.

\section{Improvements of Numerical Efficiency }
The algorithmic implementations of this study involved numerically challenging operations, %such as matrix square root or NLP optimizations. 
whose efficiency could be improved through modifications described in the following. The code was implemented in Matlab (Version R2022b) using the System Identification Toolbox (Version 10.0) \citep{TheMathWorksInc..2022}.

\subsection{Cholesky Decomposition}
During each iteration of the UKF, the matrix square root of $P_k$ needs to be computed, which is typically done by means of the Cholesky decomposition \citep{vanderMerwe.2004b}. 
The conventional Matlab command \code{chol} returns Cholesky factors of $P_k$ but requires it to be positive definite. In theory, $P_k$ is always positive definite provided $P_0$ was chosen positive definite. However, due to numerical inaccuracies such as round-off and truncation errors, $P_k$ can lose its positive definiteness \citep{Holmes.2009}. For this reason, \citep{Hartikainen.2020} developed the modified command \code{schol}, which also allows positive semi-definite matrices $P_k$. Throughout all implementations of this study, \code{schol} was used.
\subsection{Square Root Version} 
The computational effort associated with computing the Cholesky factors can be reduced by directly updating the square root of $P_k$ in each iteration. Therefore, the square root UKF was proposed in \citep{vanderMerwe.2004b} for additive noise and is reported to show improved numerical stability compared with the conventional UKF \citep{Vivo.2017}. 
%Instead of computing square root of $P_k$ in every iteration, update the square root directly. For increased numerical efficiency, QR-decomposition is applied. 
% 
\subsection{Accelerating Optimization Through Gradients and Hessian}
In Matlab, the standard solver for NLPs is \code{fmincon}, which by default approximates gradient and Hessian of the cost function through finite differences \citep{TheMathWorksInc..2023b}. When providing analytic expressions of gradient and Hessian, computational efficiency as well as numerical robustness can be vastly increased. For this reason, these expressions are derived in the following. 

The inequality constraints are merged into the cost function through Lagrange multipliers $\mu$, delivering the Lagrangian
\begin{equation}
\label{eq:meth:Lagrangian}
L_{k,i} = J_{k,i}^{NLP} + \mu^T (A \chi_{k,i}^x - b)
\end{equation}
with $\mu \in \mathbb{R}^{m \times 1}$. The gradient of the Lagrangian reads 
\begin{equation}
    \frac{d}{d \chi_{k,i}^x} L_{k,i} = \frac{d}{\chi_{k,i}^x} J_{k,i}^{NLP} + \mu^T A. 
\end{equation}
Further, the gradient of the cost function is computed as 
\begin{subequations}
\label{eq:meth:gradient}
%\begin{IEEEeqnarray}{ll}
\begin{align}
\frac{d}{\chi_{k,i}^x} J_{k,i}^{NLP} &= \frac{\partial J_{k,i}^{NLP}}{\partial h} \frac{\partial h}{\partial \chi_{k,i}^x} + \frac{\partial J_{k,i}^{NLP}}{\partial \chi_{k,i}^x} , \quad \text{where} \\
\frac{\partial J_{k,i}^{NLP}}{\partial h} &= -2\left(y - h(\chi_{k,i}^x)\right)^T R^{-1} \quad \text{and}\\
\frac{\partial J_{k,i}^{NLP}}{\partial \chi_{k,i}^x} &= 2\left( \chi_{k,i}^x - \chi_{k,i}^{x-} \right)^T \left(P_k^-\right)^{-1}.
\end{align}
%\end{IEEEeqnarray}
\end{subequations}
with
\begin{equation}
\frac{\partial h}{\partial \chi_{k,i}^x} = \left. \frac{\partial h}{\partial x} \right|_{\chi_{k,i}^x}.
\label{eq:meth:gradient_dhdx_sigma}
\end{equation}
For linear output equations as in \eqref{eq:meth:linearOutput}, \eqref{eq:meth:gradient_dhdx_sigma} reduces to $C_k$. Finally, the Hessian of the Lagrangian reads 
\begin{equation}
\label{eq:meth:hessian}
\frac{d^2}{d (\chi_{k,i}^x)^2} L_{k,i} = 2 \left(P_k^-\right)^{-1} + 2 \left( R_k^{-1} \frac{\partial h}{\partial \chi_{k,i}^x}\right)^T \frac{\partial h}{\partial \chi_{k,i}^x}.
\end{equation}
\section{Modelling of the Anaerobic Digestion Process}
\label{sec:meth:ADModels}
The model equations are derived from the ADM1-R4 proposed by \citep{Weinrich2021b}. Water and nitrogen were omitted because they are quasi-autonomous states as shown in \citep{Hellmann.2023b}. Furthermore, the gas phase was neglected to describe only the core of AD process, that is the degradation from macro nutrients to dissolved methane (\chfour) and carbon dioxide (\cotwo). 

The state vector comprises the mass concentrations (in \si{\kilogram\per\cubic\meter}) of the six states \chfour, \cotwo, carbohydrates (ch), proteins (pr), lipids (li) and microbial biomass (bac)
\begin{equation*}
\label{eq:meth:state}
x = [x_1 \, x_2 \, x_3 \, x_4 \, x_5 \, x_6 ]^T = [S_\mathrm{ch4} \, S_\mathrm{co2} \, X_\mathrm{ch} \, X_\mathrm{pr} \, X_\mathrm{li} \, X_\mathrm{bac} ]^T.
\end{equation*}
The state differential equations read as follows: 
\begin{subequations} \label{eq:meth:ADM1R4Core_x}
\begin{align}
\dot x_1 &= c_1 \left(\xi_1 - x_1\right) u + a_{11} c_2 x_3 + a_{12} c_3 x_4 + a_{13} c_4 x_5 \\
\dot x_2 &= c_1 \left(\xi_2 - x_2\right) u + a_{21} c_2 x_3 + a_{22} c_3 x_4 + a_{23} c_4 x_5 \\
\dot x_3 &= c_1 \left(\xi_3 - x_3\right) u - c_2 x_3 + a_{34} c_5 x_6 \\
\dot x_4 &= c_1 \left(\xi_4 - x_4\right) u - c_3 x_4 + a_{44} c_5 x_6 \\ 
\dot x_5 &= c_1 \left(\xi_5 - x_5\right) u - c_4 x_5 + a_{54} c_5 x_6\\ 
\begin{split}
\dot x_6 &= c_1 \left(\xi_6 - x_6\right) u + a_{61} c_2 x_3 + a_{62} c_3 x_4 + 
\\
&\hspace{12pt} + a_{63} c_4 x_5 - c_5 x_6. 
\end{split}
\end{align}
\end{subequations}
The dissolved gas concentrations of \chfour and \cotwo as well as microbial biomass were assumed to be measurable:
\begin{equation}
\label{eq:meth:ADM1R4Core_y}
y = \begin{bmatrix} x_1 \\ x_2 \\ x_6 \end{bmatrix} = 
\begin{bmatrix}
1 & 0 & 0 & 0 & 0 & 0 \\
0 & 1 & 0 & 0 & 0 & 0 \\
0 & 0 & 0 & 0 & 0 & 1 \end{bmatrix} x. 
\end{equation}
The substrate feed volume flow acts as the control variable $u$. Model parameters $a$, $c$ and $\xi$ were derived as summarized in the appendix. The simulation scenario and the model are also described in more detail there. The solver \code{ode15s} was used to minimize discretization errors through a variable step size.
\subsection{Simulation Scenario}
\label{sec:res:simulation-scenario}
A pilot-scale AD reactor with \SI{100}{\liter} liquid volume was fed dynamically for one week with a substrate mix of maize silage and cattle manure, starting in steady state conditions (with normalized volatile solids loading rate of \SI{6.6}{\kilogram\per\cubic\meter\day} and retention time of \SI{4.5}{d}). The feeding pattern is specified in Table~\ref{tab:meth:feedProperties}.
\begin{table}[hbt]
	\renewcommand{\arraystretch}{1.1}
	\caption{Properties of two substrate feeding peaks during simulation scenario (otherwise no feeding).}
	\begin{center}
		\begin{tabular}{|c|c|c|c|}
			\hline
			%			\textbf{Table}&\multicolumn{3}{|c|}{\textbf{Table Column Head}} \\
			%			\cline{2-4} 
			%			\textbf{Head} & \textbf{\textit{Table column subhead}}& \textbf{\textit{Subhead}}& \textbf{\textit{Subhead}} \\
			% 
			\textbf{Feeding no.} & \textbf{Volume flow [\si{\cubic\meter\per\day}]} & \textbf{Start [\si{\day}]} & \textbf{Duration [\si{\day}]} \\
			\hline
			1 & 168 & 2.5 & 0.5\\
			\hline 
			2 & 72 & 5.5 & 1 \\
			\hline
			% \multicolumn{5}{l}{bla $^{\mathrm{a}}$average value of}\\
		\end{tabular}
		\label{tab:meth:feedProperties}
	\end{center}
\end{table}
Measurements were assumed to be taken every $\Delta t=\SI{0.5}{\hour}$, resulting in $N=337$ samples. Nominal measurements were superimposed with additive, zero-mean Gaussian noise with standard deviations $\sigma_i$ as described in the appendix. 

To ensure comparability among all implemented UKFs, all of them were equally tuned as follows \citep{Schneider.2013}:
%
%\begin{IEEEeqnarray*}{ll}
\begin{align}
x_0 &= [4.09, \, 10.52, \, 11.04, \, 2.57, \, 0.96, \, 2.02]^T \\
\hat x_0 &= [2.20, \, 19.30, \, 24.94, \, 2.22, \, 0.31, \, 2.64]^T \\
P_0 &= \mathrm{diag}\{\left(\hat x_0 - x_0 \right)^2\} \\
R &= 1.5\cdot \mathrm{diag}\{\left(\sigma_i\right)^2\} \\
Q &= \mathrm{diag}\{[1, \, 1, \, 1, \, 1, \, 1, \, 1]\}.
\end{align}
%\end{IEEEeqnarray*}
%  
A plant-model mismatch as stated in Table~\ref{tab:meth:plant-model-mismatch} was assumed. 
\begin{table}[bt]
	\renewcommand{\arraystretch}{1.1}
	\caption{Model parameters used for synthetic measurement creation (true value) and for unscented Kalman filtering (UKF value). More digits for the UKF values are stated in Table~\ref{tab:app:R4-Core-nomenclature}.}
	\begin{center}
		\begin{tabular}{|l|l|l|l|l|}
			\hline
%			\textbf{Table}&\multicolumn{3}{|c|}{\textbf{Table Column Head}} \\
%			\cline{2-4} 
%			\textbf{Head} & \textbf{\textit{Table column subhead}}& \textbf{\textit{Subhead}}& \textbf{\textit{Subhead}} \\
% 
			& $\mathbf{c_2}$ \textbf{[\si{\per\hour}]} & $\mathbf{c_3}$ \textbf{[\si{\per\hour}]} & $\mathbf{c_4}$ \textbf{[\si{\per\hour}]} & $\mathbf{c_5}$ \textbf{[\si{\per\hour}]} \\
			\hline
            true value & 0.25 & 0.20 & 0.10 & 0.020 \\
            \hline
            UKF value & 0.32 & 0.26 & 0.13 & 0.026 \\
            \hline
			% \multicolumn{5}{l}{bla $^{\mathrm{a}}$average value of}\\
		\end{tabular}
		\label{tab:meth:plant-model-mismatch}
	\end{center}
\end{table}
\subsection{Normalized Root Mean Squared Error}
To evaluate estimation accuracy, the normalized root mean squared error (NRMSE) between estimated and true values was used:
\begin{equation}
\label{eq:meth:nRMSE}
\mathrm{NRMSE} = \frac{\sqrt{1/N\sum_i \left( \hat{x}_i - x_i \right)^2}}{1/N\sum_i x_i}.
\end{equation}
%
%It is computed for each state individually and then averaged over all measurable and non-measurable states
% Throughout this study, nRMSE values are stated as the average over 20 runs.
This ratio can be computed both for states $x$ and outputs $y$, and is denoted accordingly as NRMSE\textsubscript{x} or NRMSE\textsubscript{y}. 

\section{Simulation studies}
Manifold UKF variants have been implemented as summarized in Table~\ref{tab:meth:overviewImplementations}. 
\begin{table}[bt]
	\renewcommand{\arraystretch}{1.1}
	\caption{Overview of all implemented UKF versions with short description. Explanations are given in the text.}
	\begin{center}
		\begin{tabular}{|l||l|l||l|}
			\hline
%			\textbf{Table}&\multicolumn{3}{|c|}{\textbf{Table Column Head}} \\
%			\cline{2-4} 
%			\textbf{Head} & \textbf{\textit{Table column subhead}}& \textbf{\textit{Subhead}}& \textbf{\textit{Subhead}} \\
% 
			\textbf{Description} & \multicolumn{2}{c||}{\textbf{Unconstrained}} & \textbf{Constrained} \\
			\hline
            Toolbox$^{\mathrm{a}}$ & \multicolumn{2}{c||}{UKF-sysID} & \\
            \hline
            Square root & UKF-SR & UKF-SR-$\gamma$ & \\
            \hline
            Additive & UKF-add & UKF-add-$\gamma$ & cUKF-add \\
            \hline
            Augmented & UKF-aug & UKF-aug-$\gamma$ & cUKF-aug \\
            \hline
            Fully augm.$^{\mathrm{b}}$ & UKF-fully-aug & UKF-fully-aug-$\gamma$ & cUKF-fully-aug \\
            \hline
			\multicolumn{4}{l}{$^{\mathrm{a}}$ Matlab System Identification Toolbox, $^{\mathrm{b}}$ fully augmented}\\
		\end{tabular}
		\label{tab:meth:overviewImplementations}
	\end{center}
\end{table}
They are classified as unconstrained and constrained UKFs. The former are described in Section~\ref{sec:res:unconstrained}, the latter in Section~\ref{sec:res:constrained}. Section~\ref{sec:res:summary} summarizes the performance of all implemented UKF variants.
\subsection{Unconstrained Case}
\label{sec:res:unconstrained}
Unconstrained UKF implementations using the default sigma point scaling are discussed first. Then the effect of reduced sigma point scaling is analyzed. UKF-sysID denotes the UKF implemented with the Matlab System Identification Toolbox assuming additive noise. It serves as a benchmark.
\subsubsection{Nominal Sigma Point Scaling}
\label{sec:res:UKF}
The following algorithms were implemented and compared with the benchmark \mbox{UKF-sysID}, all 
considering nominal sigma point scaling \eqref{eq:meth:UKF-scaling} with default UKF tuning parameters \eqref{eq:meth:defaultTuning}, see Table~\ref{tab:meth:overviewImplementations}: square root UKF according to \citep{vanderMerwe.2001} (UKF-SR), as well as the unconstrained UKFs according to \citep{Kolas.2009} assuming additive noise (\mbox{UKF-add}), non-additive process noise (UKF-aug) and non-additive process and measurement noise (UKF-fully-aug). Note that although the model used in this study assumes additive process and measurement noise, the augmented and fully augmented algorithm variants can still be applied.

\begin{figure}[bt]
	\begin{center}
		\includegraphics[width=0.8\linewidth]{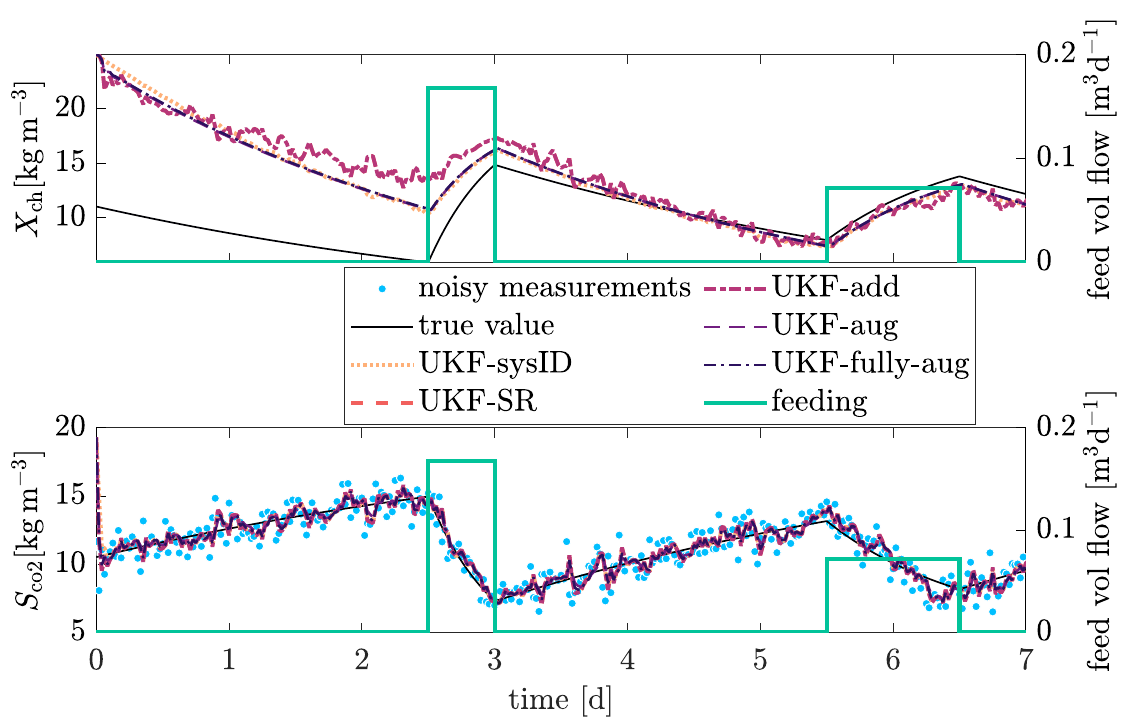}
		\caption{Comparison of state estimation quality through different UKFs with nominal sigma point scaling. Top: Concentration of carbohydrates and corresponding estimations. Bottom: Concentration of dissolved CO\textsubscript{2}, measurements and corresponding estimations.}
		\label{fig:res:UKF}
	\end{center}
\end{figure}

Figure~\ref{fig:res:UKF} compares state estimation performance of the unconstrained UKF variants by means of carbohydrates and \cotwo concentrations.
The simulation model \eqref{eq:meth:ADM1R4Core_x}, \eqref{eq:meth:ADM1R4Core_y} contains three measurable and three non-measurable states. However, qualitative filter performance for all measurable and all non-measurable states is largely the same. This holds for all implemented algorithms. Therefore, in the following only one each is shown, which is largely representative of the corresponding two other ones, exceptions are highlighted.

As the non-measurable state, the carbohydrate concentration is chosen 
%over proteins and lipids 
for being the largest nutrient fraction. % and thus expressing the largest numbers. 
As a measurable state, dissolved carbon dioxide concentration is chosen %over dissolved methane and biomass 
since it exhibits the largest values and is the easiest one to measure in reality \citep{Neddermeyer.2015}.
%in laboratory setups it can be measured with relatively simple equipment, see e.g. \citep{Neddermayer14.8.2020}.

Smoothing of measurable states, such as dissolved \cotwo in the bottom of Figure~\ref{fig:res:UKF}, is very similar for all UKF versions. %, the graphs in the bottom of Figure~\ref{fig:res:UKF} almost overlap. 
This is reinforced through nearly identical values of NRMSE\textsubscript{y}, see Table~\ref{tab:res:UKF}. The noise-free output is not met exactly, but the filters clearly smooth the noisy measurements, underlined by low NRMSE\textsubscript{y}. 

\begin{table}[bt]
	\renewcommand{\arraystretch}{1.1}
	\caption{Comparison of unconstrained UKF performance with default sigma point scaling ($\gamma=2.4495$) according to \eqref{eq:meth:UKF-scaling} and \eqref{eq:meth:defaultTuning}.}
	\begin{center}
		\begin{tabular}{|l|l|l|l|}
			\hline
%			\textbf{Table}&\multicolumn{3}{|c|}{\textbf{Table Column Head}} \\
%			\cline{2-4} 
%			\textbf{Head} & \textbf{\textit{Table column subhead}}& \textbf{\textit{Subhead}}& \textbf{\textit{Subhead}} \\
			\textbf{Algorithm} & \textbf{NRMSE\textsubscript{x}}$^{\mathrm{a}}$ & \textbf{NRMSE\textsubscript{y}}$^{\mathrm{b}}$ & \textbf{Run time [s]} \\
			\hline
            UKF-sysID & 0.4888 & 0.1152 & 1.73 \\
            \hline
            UKF-SR & 0.8533 & 0.1157 & 2.16 \\
            \hline 
            UKF-add & 0.8533 & 0.1157 & 2.29 \\
            \hline 
            UKF-aug & 0.3599 & 0.0934 & 3.66 \\
            \hline 
            UKF-fully-aug & 0.3599 & 0.1081 & 4.14 \\
			\hline 
            \multicolumn{4}{l}{$^{\mathrm{a}}$average value of all non-measurable states $^{\mathrm{b}}$average value of all}\\
            \multicolumn{4}{l}{measurable states}\\
            \end{tabular}
		\label{tab:res:UKF}
	\end{center}
\end{table}
For the non-measurable states, such as carbohydrates in the top of Figure~\ref{fig:res:UKF}, all filters approach the true trajectory despite the initial estimation error. As of $t\approx \SI{4}{d}$, estimations agree well with true values. In light of a plant-model mismatch, it is plausible that they do not match exactly. 

Yet, convergence behavior of the algorithms differs. \mbox{UKF-add} and UKF-SR deliver identical results, and hence show overlapping graphs and the same NRMSE\textsubscript{x}. They both do not involve augmentation, so essentially their flow of algorithmic operations is the same. Their identical performance emphasizes that the square root formulation of the additive UKF derived in \citep{vanderMerwe.2001} is equivalent to the conventional additive UKF. However, both UKF-add and UKF-SR show a %higher discrepancy from the true value 
slower convergence than UKF-sysID, clearly visible before the first feeding, and reflected in a higher NRMSE\textsubscript{x}. Since UKF-sysID and UKF-SR are both based on the square root UKF \citep{TheMathWorksInc..2023}, their graphs should match. However, they may deviate because UKF-sysID internally uses slightly different sigma point scaling and weighting than stated in \citep{vanderMerwe.2001}. Another reason might be different numerical performance. This is likely the case, especially because all three additive UKFs delivered negative state estimates for the lipids concentration ($x_5$) before the first feeding. To this end, UKF-add exhibited the lowest estimates of $x_5$. Since negative concentrations are beyond the physically meaningful domain of the model, these negative estimates of $x_5$ might also influence the other state estimates. 
%The the algorithmic details in the code of UKF-sysID and UKF-SR 
%there exist manifold ways to hard-code the measurement update step especially for the square-root version, which can result in different numerical behavior and is assumed to cause the discrepancy here. 

By contrast, the augmented versions UKF-aug and \mbox{UKF-fully-aug} deliver no negative and much less noisy state estimates, see Figure~\ref{fig:res:UKF}. They approach the true trajectory faster and yield a lower NRMSE\textsubscript{x} than UKF-sysID, see Table~\ref{tab:res:UKF}. Run times of all additive UKF versions over the entire simulation horizon are in the same range of about \SI{2}{\second}. Among them, UKF-sysID is the fastest. This is plausible since it is a commercial implementation optimized for numerical efficiency. By comparison, run times of UKF-aug and UKF-fully-aug are clearly higher, reflecting the higher resulting system order, cf. \eqref{eq:meth:augmentation} and \eqref{eq:meth:fulAugmentation}. 
%
%chol, cholupdate, sigma point selection (lower triangular or upper triangular), qr $\rightarrow$ can't look into matlabs code.
%
% UKF-add also delivers exact same P-matrix as according to \citep{vanderMerwe.2001} with max. difference in P-matrix in the range of 1E-14 and 1E-16. overall, UKF-sysID exhibits less noisy estimation of non-measurable state. 
%
\subsubsection{Reduced Sigma Point Scaling}
\begin{figure}[tb]
	\begin{center}
		\includegraphics[width=0.8\linewidth]{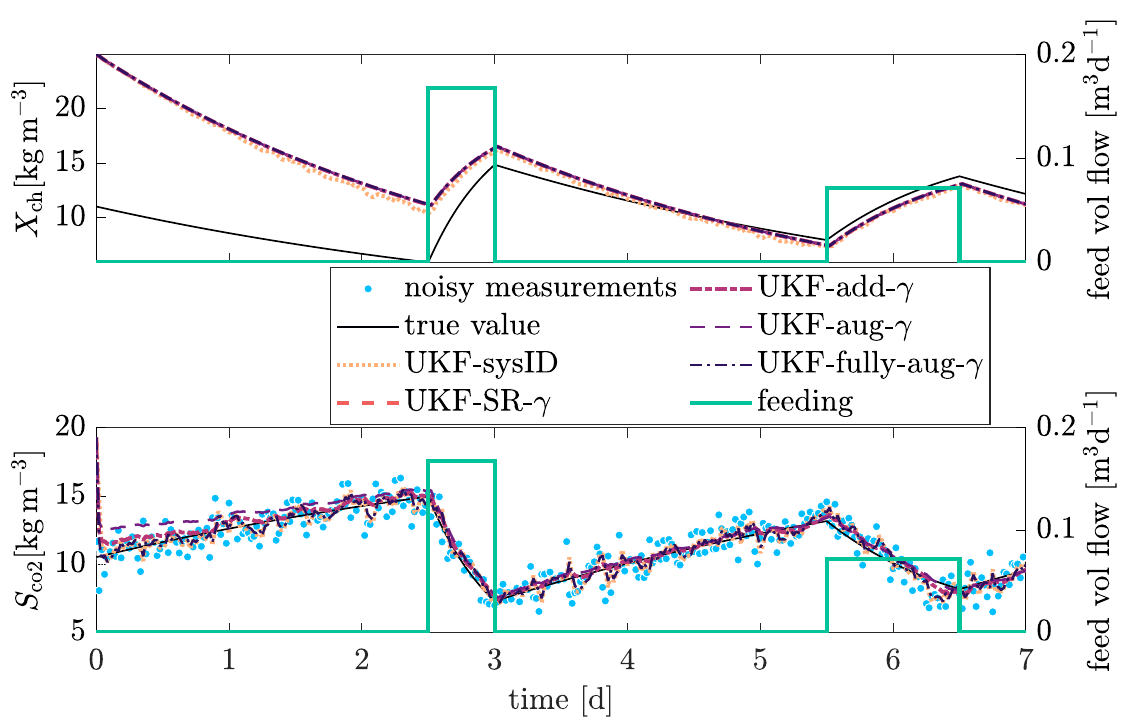}
		\caption{Comparison of state estimation quality through different UKFs with modified sigma point scaling (reduced scaling factor $\gamma=1$). For the benchmark UKF-sysID, conventional tuning was retained for comparison. Top: Concentration of carbohydrates and corresponding estimations. Bottom: Concentration of dissolved CO\textsubscript{2}, measurements and corresponding estimations.}
		\label{fig:res:UKF-gamma}
	\end{center}
\end{figure}

For all unconstrained UKF variants, except the benchmark, the sigma point scaling was reduced by changing the scaling factor $\gamma$ from 2.4495 (according to nominal UKF scaling \eqref{eq:meth:UKF-scaling}) to 1, though without changing the weighting in \eqref{eq:meth:UKF-weights}. The resulting performance is illustrated in Figure~\ref{fig:res:UKF-gamma}, with corresponding graphs indicated by the name extension -$\gamma$. 
\begin{table}[tb]
	\renewcommand{\arraystretch}{1.1}
	\caption{Comparison of unconstrained UKF performance with modified sigma point scaling ($\gamma=1$).}
	\begin{center}
		\begin{tabular}{|l|l|l|l|}
			\hline
%			\textbf{Table}&\multicolumn{3}{|c|}{\textbf{Table Column Head}} \\
%			\cline{2-4} 
%			\textbf{Head} & \textbf{\textit{Table column subhead}}& \textbf{\textit{Subhead}}& \textbf{\textit{Subhead}} \\
			\textbf{Algorithm} & \textbf{NRMSE\textsubscript{x}}$^{\mathrm{a}}$ & \textbf{NRMSE\textsubscript{y}}$^{\mathrm{b}}$ & \textbf{Run time [s]} \\
			\hline
        	UKF-SR-$\gamma$ & 0.3733 & 0.0657 & 2.18 \\
        	\hline 
        	UKF-add-$\gamma$ & 0.3733 & 0.0657 & 2.15 \\
        	\hline 
        	UKF-aug-$\gamma$ & 0.3691 & 0.0647 & 3.84 \\
        	\hline 
        	UKF-fully-aug-$\gamma$ & 0.3695 & 0.1046 & 4.40 \\
			\hline 
            \multicolumn{4}{l}{$^{\mathrm{a}}$average value of all non-measurable states $^{\mathrm{b}}$average value of all}\\
            \multicolumn{4}{l}{measurable states}\\
            \end{tabular}
		\label{tab:res:UKF-gamma}
	\end{center}
\end{table}
Performance improves especially for the additive variants UKF-add-$\gamma$ and \mbox{UKF-SR-$\gamma$}. This manifests in lower NRMSE values than for nominal scaling and also than \mbox{UKF-sysID}. At the same time, about the same run times are maintained, as illustrated in Tables~\ref{tab:res:UKF} and \ref{tab:res:UKF-gamma}. Moreover, no negative values for estimated lipids concentrations ($x_5$) are obtained anymore. 

For the augmented versions, the positive effect of reducing $\gamma$ is not as clear: NRMSE\textsubscript{y} reduces for \mbox{UKF-aug-$\gamma$}, whereas NRMSE\textsubscript{x} slightly increases. For the fully augmented version both NRMSE values remain almost unchanged at an acceptable level. We conclude that for the given model and simulation scenario, sigma point scaling according to \citep{vanderMerwe.2004b} does not necessarily deliver the best possible estimation performance. This especially holds for the additive noise case. 

Note that reducing $\gamma$ does not affect the aggregation step of the scaled unscented transformation, since it is still holds that $\sum_i W_i^x=1$, cf. \eqref{eq:meth:UKF-weights}. Moreover, reducing the scaling factor $\gamma$ could also be achieved through negative values of $\kappa$, see \eqref{eq:meth:UKF-scaling}. However, this affects the parameter $\lambda$ and hence the weights of the mean and covariance. The scaled unscented transformation eventually still remains unchanged, although scaled linearly in $\lambda$. Choosing negative values of $\kappa$ delivered exactly the same results as those obtained with $\kappa=0$. We thus conclude that for the purpose of smoothing noisy measurements, reducing the scaling of sigma points can be beneficial to estimation performance.
\subsection{Constrained Case}
\label{sec:res:constrained}
The algorithms presented so far did not explicitly account for state constraints. \citep{Kolas.2009} suggested to introduce clipping in various locations of the unconstrained UKFs to address state estimates beyond physically meaningful bounds. 
%In the biological model of this study, %\eqref{eq:meth:ADM1R4Core_x}, \eqref{eq:meth:ADM1R4Core_y}
%states represent concentrations which must be non-negative to be physically meaningful. 
However, clipping diminished the estimation quality in our case (results not shown). This behavior appears to be reasonable: abrupt clipping without simultaneously adapting the sigma point distribution distorts the unscented transformation. A remedy may lie in applying the truncation method \citep{Simon.2010} proposed for linear Kalman filters and extending it for UKFs. This was, however, not further pursued here.

By contrast, accounting for state constraints through solving the optimization problem \eqref{eq:meth:optimizationProblem} could be shown to improve estimation performance especially for additive noise, which is explained in the following.

\begin{figure}[bt]
	\begin{center}
		\includegraphics[width=0.8\linewidth]{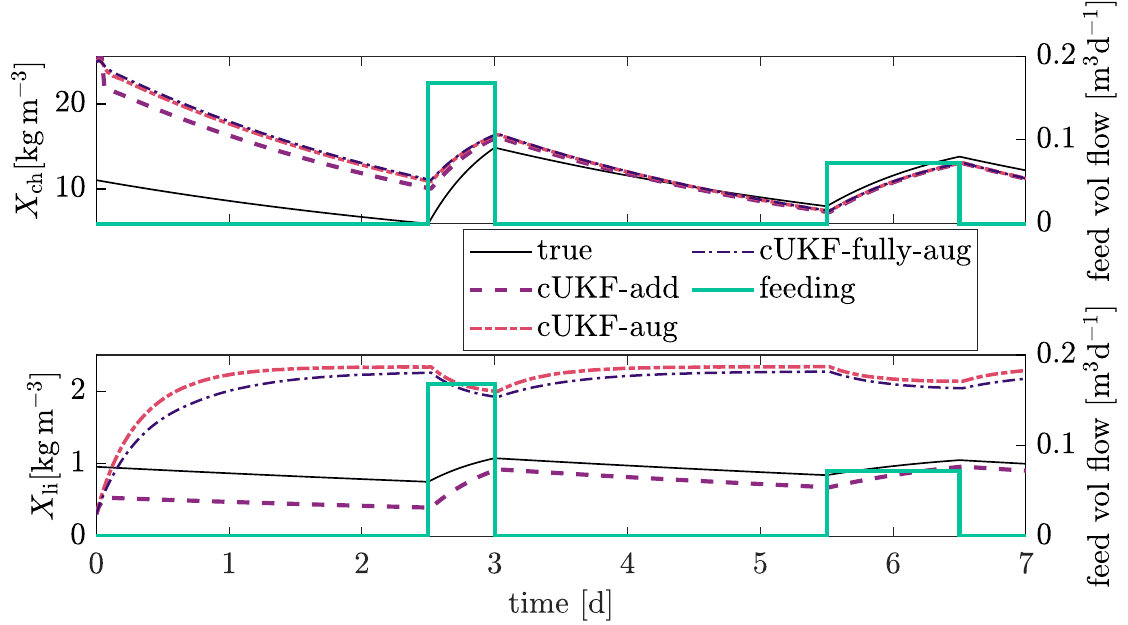}
		\caption{Comparison of state estimation quality through different constrained UKFs. Top: Concentration of carbohydrates and corresponding estimations. Bottom: Concentration of lipids and corresponding estimations.}
		\label{fig:res:cUKF}
	\end{center}
\end{figure}

The constrained UKFs were implemented for all three noise cases, delivering additive (cUKF-add), augmented (\mbox{cUKF-aug}) and fully augmented cUKFs (cUKF-fully-aug). Note that nominal sigma point scaling \eqref{eq:meth:UKF-scaling} was applied. Furthermore, the optimization of each cUKF could be described by either the NLP \eqref{eq:meth:costFunctionNLP} or the QP formulation \eqref{eq:meth:costFunctionQP} since the model output \eqref{eq:meth:ADM1R4Core_y} is linear \citep{Kolas.2009}. For the NLP, gradients and the Hessian were either approximated by finite differences (\mbox{cUKF-NLP}); by providing analytic expressions for the gradients (cUKF-NLP-grad); or both gradients and Hessian (cUKF-NLP-grad-hess). All setups of the optimization problems for a given noise case delivered the same estimations. Therefore, Figure~\ref{fig:res:cUKF} shows the cUKF performances only with respect to a given noise case. %and not for a specific optimization setup. 
Furthermore, Table~\ref{tab:res:cUKF} summarizes corresponding NRMSE values. The stated run times apply for the NLP setup with finite differences.

The measurable states are smoothed comparably well as for the unconstrained UKFs, mirrored by NRMSE\textsubscript{y} values in the same range as for nominal and reduced sigma point scaling, cf. Tables~\ref{tab:res:UKF}, \ref{tab:res:UKF-gamma} and \ref{tab:res:cUKF}. For this reason, the graphs of \cotwo measurements and smoothed estimates are not shown here again. 
\begin{table}[bt]
	\renewcommand{\arraystretch}{1.1}
	\caption{Comparison of constrained UKF performance. Run times apply for NLP formulation without gradients and Hessian.}
	\begin{center}
		\begin{tabular}{|l|l|l|l|}
			\hline
%			\textbf{Table}&\multicolumn{3}{|c|}{\textbf{Table Column Head}} \\
%			\cline{2-4} 
%			\textbf{Head} & \textbf{\textit{Table column subhead}}& \textbf{\textit{Subhead}}& \textbf{\textit{Subhead}} \\
			\textbf{Algorithm} & \textbf{NRMSE\textsubscript{x}}$^{\mathrm{a}}$ & \textbf{NRMSE\textsubscript{y}}$^{\mathrm{b}}$ & \textbf{Run time [s]} \\
			\hline
            cUKF-add & 0.2897 & 0.1040 & 54.17 \\
            \hline 
            cUKF-aug & 0.6746 & 0.0902 & 129.57 \\
            \hline 
            cUKF-fully-aug & 0.6345 & 0.0843 & 158.06 \\
			\hline 
            \multicolumn{4}{l}{$^{\mathrm{a}}$average value of all non-measurable states $^{\mathrm{b}}$average value of all}\\
            \multicolumn{4}{l}{measurable states}\\
		\end{tabular}
		\label{tab:res:cUKF}
	\end{center}
\end{table}

Instead, Figure~\ref{fig:res:cUKF} illustrates estimation performance for the (non-measurable) carbohydrates and lipids concentrations $X_\mathrm{ch}$ and $X_\mathrm{li}$. The latter exhibited negative values for all additive, unconstrained UKFs with nominal values of $\gamma$, see Section~\ref{sec:res:UKF}.
It is clear that for carbohydrates, estimates of all three algorithms do not differ much, and they all approach the true trajectory well. This is reflected in very similar values of NRMSE\textsubscript{x} for carbohydrates between 0.024 and 0.029. 
The augmented versions cUKF-aug and \mbox{cUKF-fully-aug} are slightly closer to the true value than cUKF-add.

By contrast, the lipids estimations differ significantly, \mbox{Figure}~\ref{fig:res:cUKF} bottom. This might, on the one hand, be attributed to the lower order of magnitude of $X_\mathrm{li}$. % and its small initial value.
\citep{Vachhani.2006} mentioned that for estimates close to the constraints (i.e. very low concentrations in our case), the optimization might result in a biased estimate. On the other hand, closely inspecting the algorithms of \mbox{cUKF-aug} and cUKF-fully-aug reveals a crucial aspect not addressed in \citep{Kolas.2009}. Therein the authors adopt the NLP from \citep{Vachhani.2006} and apply it to the augmented cases, while in \citep{Vachhani.2006} it was formulated for the additive noise case. To this end, it remains unclear how cUKF-aug and cUKF-fully-aug effectively differ from each other aside from the augmentation, since the intermediate steps for computing the estimated output $\hat y$ and the corresponding covariance matrix $P_{y_k y_k}$ do not come into effect in the constrained case. Instead, the outputs resulting from the updated sigma points $h(\chi_{k,i}^x)$ are computed in each iteration of the optimization, \eqref{eq:meth:costFunctionNLP}. Additionally, \citep{Kolas.2009} apply the nominal output $h(x_k)$ in the cost function of the fully augmented noise case, neglecting the non-additive formulation $h(x_k,v_k)$ used in the unconstrained UKFs. 

We thus conclude that the NLP formulation \eqref{eq:meth:costFunctionNLP} might only deliver reliable state estimates for the additive noise case (cUKF-add) for which it was originally proposed, especially for state estimates close to the bounds. This may also explain why the graphs of cUKF-aug and cUKF-fully-aug both show similarly poor lipids estimates in Figure~\ref{fig:res:cUKF}. The poor estimates of $X_\mathrm{li}$ through cUKF-aug and cUKF-fully-aug also dominate the comparably high values of NRMSE\textsubscript{x} %representing the average of all non-measurable state estimates
in Table~\ref{tab:res:cUKF}. 

% cUKF-NLP denotes the constrained UKF applying the NLP cost function \eqref{eq:meth:costFunctionNLP}, solved through Matlab's fmincon solver. For cUKF-QP, on the other hand, the QP formulation \eqref{eq:meth:costFunctionQP} was applied and solved through Matlab's quadprog solver. As explained by \citep{Kolas.2009}, both formulations are equivalent in the case of linear output equations which applies for the model of this study. Accordingly, the curves overlap and NRMSE values are identical.

Estimation results of all optimization setups for a given noise case are identical. This emphasizes that for linear output equations the QP reformulation of the NLP by \citep{Kolas.2009} is indeed equivalent, and that the provision of gradients and Hessian increases numerical efficiency without jeopardizing accuracy. 

\begin{figure}[tb]
	\begin{center}
		\includegraphics[width=0.8\linewidth]{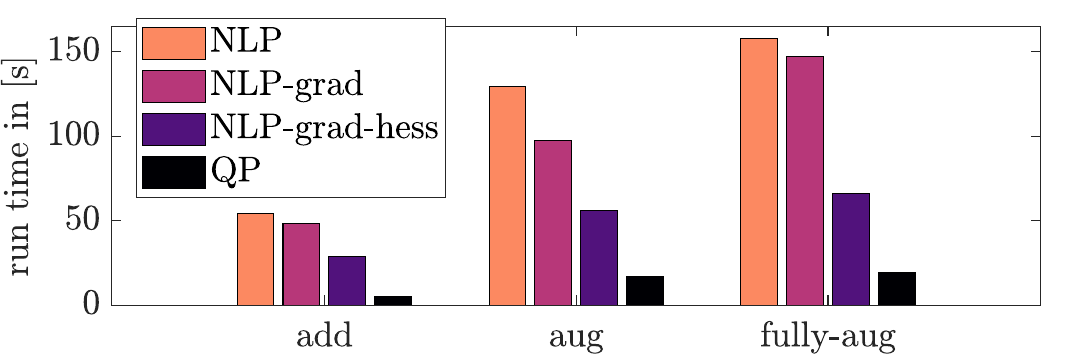}
		\caption{Run times of cUKF versions with all three noise cases in different optimization setups.}
		\label{fig:res:cUKF_tRun}
	\end{center}
\end{figure}

Run times of the constrained UKFs are generally higher than for the unconstrained versions. However, they can be vastly reduced through a) analytic expressions of gradient and Hessian for the NLP, and b) through the QP formulation in case of linear outputs, as emphasized in Figure~\ref{fig:res:cUKF_tRun}. Given the higher resulting system order caused by augmentation, it is plausible that numerical effort for cUKF-aug and \mbox{cUKF-fully-aug} is higher than for cUKF-add. In case run time is critical, a UKF formulation specifically designed for real-time applications was recently proposed by \citep{Cantelobre.2020} and might present an alternative to the UKF designs discussed in this work.

The reduced run times of the improved NLP setups can be explained when considering the average number of iterations and cost function calls before convergence, summarized in Table~\ref{tab:res:cUKF-optimizer-performance}. By default, \code{fmincon} solves the NLP with an interior-point algorithm which involves to approximate gradient and Hessian of the cost function through finite differences \citep{TheMathWorksInc..2023b}. For this purpose, the cost function during one iteration of optimization needs to be computed multiple times. %for a single current value of the design variable. 
In contrast, approximation through finite differences becomes redundant with analytic expressions of gradients and Hessian. In case only the gradients are provided, the number of iterations remains the same but the number of cost function calls drops from 104 to 19. Moreover, additionally providing the Hessian reduces the number of iterations from 14 to 5, and consequently further diminishes the number of cost function calls from 19 to 6. For cUKF-QP, the solver \code{quadprog} makes use of algorithms optimized for QPs and thus required even fewer iterations.
%
%This study involved only linear inequalities, which do not require corresponding analytic expressions for gradient and Hessian, see \eqref{eq:meth:Lagrangian} and \eqref{eq:meth:hessian}. This would of course change for nonlinear equalities and inequalities.

% XY: also covergence is enhanced in cases where optimum is close to boundary because finite differences might involve infeasible points, siehe Matlab docu!  

% XY prüfen, ob das sowohl für equalities als auch für inequalities gilt. Ja, Erstellung über struct lambda mit den feldern lambda.ineqnonlin und lambda.eqnonlin, siehe https://de.mathworks.com/help/optim/ug/writing-scalar-objective-functions.html#bu2xbye-1 

%Both numbers of function calls and iterations were averaged over one complete run with 337 samples.
%
\begin{table}[tb]
    \renewcommand{\arraystretch}{1.1}
	\caption{Comparison of optimization performance of cUKF-add for different optimization setups.}
	\begin{center}
		\begin{tabular}{|l|l|l|}
			\hline
%			\textbf{Table}&\multicolumn{3}{|c|}{\textbf{Table Column Head}} \\
%			\cline{2-4} 
%			\textbf{Head} & \textbf{\textit{Table column subhead}}& \textbf{\textit{Subhead}}& \textbf{\textit{Subhead}} \\
            \textbf{Algorithm} & \textbf{\# Function calls}$^{\mathrm{a}}$ & \textbf{\# Iterations}$^{\mathrm{b}}$ \\
			\hline
            cUKF-add-NLP & 104 & 13 \\
			\hline 
            cUKF-add-NLP-grad & 19 & 14 \\
			\hline 
            cUKF-add-NLP-grad-hess & 6 & 5 \\
            \hline
            cUKF-add-QP & - & 4 \\
            \hline
            \multicolumn{3}{l}{$^{\mathrm{a}}$median of function calls $^{\mathrm{b}}$median of iterations}
		\end{tabular}
		\label{tab:res:cUKF-optimizer-performance}
	\end{center}
\end{table}
% 
% XY: das ist falsch!
%Augmented versions of cUKF were tested in analogy to Section~\ref{sec:res:unconstrained} but resulted in poor estimation quality for non-measurable states. Moreover, reducing sigma point spread through a small value of $\gamma$ did not improve estimation quality. This underlines the conclusion that augmented versions might result in troublesome numerical behavior and additive noise variants should be preferred where possible. 
% 
\subsection{Summary}
\label{sec:res:summary}
\begin{figure}[bt]
	\begin{center}
		\includegraphics[width=0.8\linewidth]{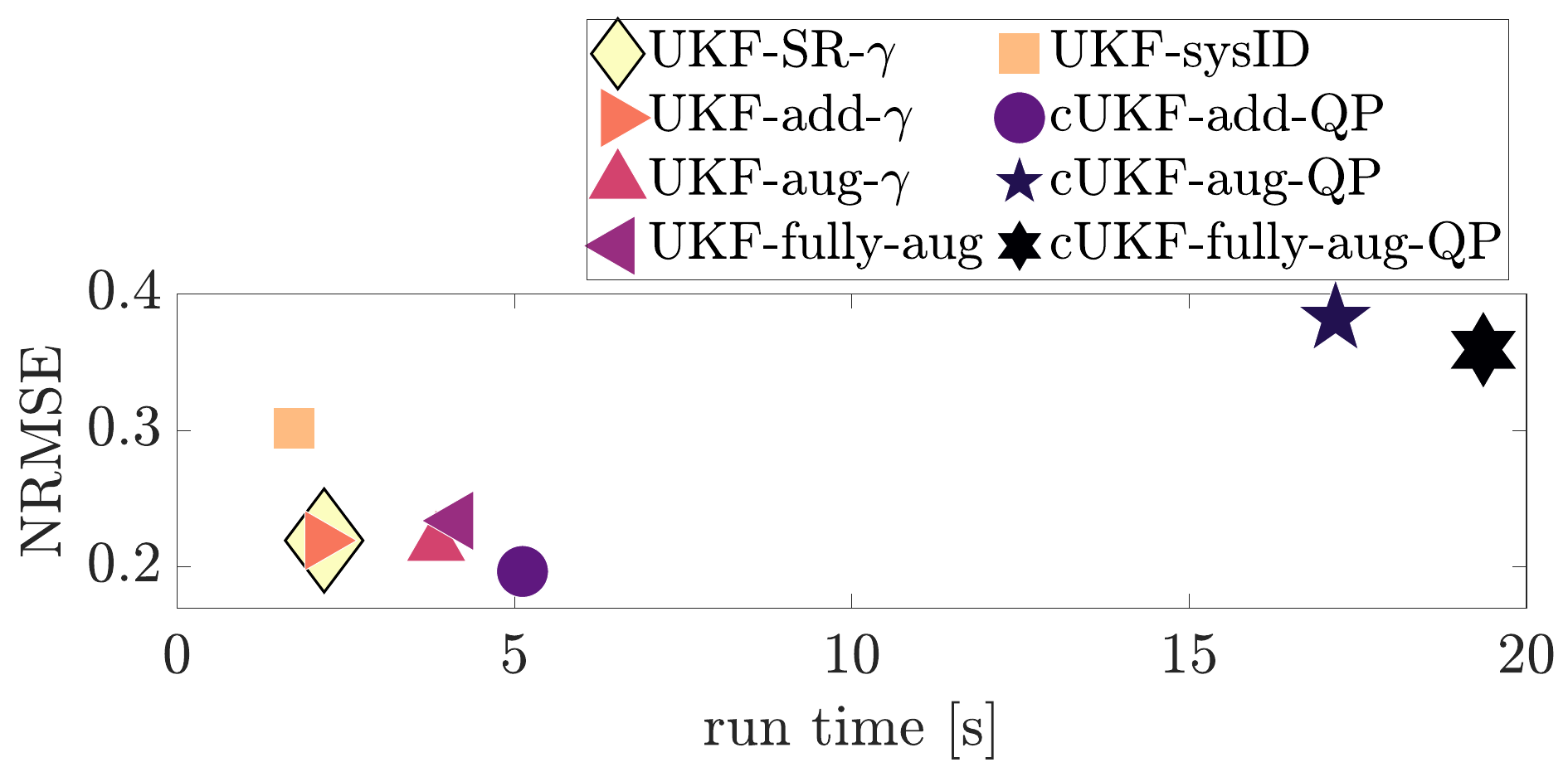}
		\caption{Comparison of best-in-class implementations of UKFs by means of run time and estimation error.}
		\label{fig:res:tCalc_vs_nRMSE}
	\end{center}
\end{figure}

Figure~\ref{fig:res:tCalc_vs_nRMSE} compares the best versions of the different classes of implemented algorithms with respect to run time and estimation accuracy, expressed as the average NRMSE over both measurable and non-measurable states. 
The unconstrained UKFs showed faster run times than the constrained UKFs. The best accuracy was achieved with cUKF-add. However, the unconstrained UKFs UKF-aug-$\gamma$ as well as the equivalent \mbox{UKF-add-$\gamma$} and UKF-SR-$\gamma$ deliver almost the same accuracy with lower run times. Among the constrained UKFs, the augmented cUKF versions cUKF-aug and cUKF-fully-aug could not compete with the additive variant cUKF-add with respect to both run time and accuracy.

\section{Conclusion}
This study examined various UKF implementations and applied them to a simplified AD model. %featuring additive noise. 
Crucial details of the underlying algorithms as well as modifications to leverage improved numerical performance were addressed. This provides useful hints for practitioners working with Kalman filters. 

The best estimation accuracy was achieved with \mbox{cUKF-add}. The QP reformulation of the underlying NLP massively reduced the run time without compromising estimation accuracy. However, it is only applicable for models with linear output equations. For models involving nonlinear output equations, cUKF-NLP-add-grad-hess presents a useful alternative although the associated run time is much higher, see Figure~\ref{fig:res:cUKF_tRun}.
%The necessary analytic expressions of gradients and Hessian can be obtained through symbolic calculation, requiring only a few lines of code. 

If run time is critical, the unconstrained variants UKF-SR-$\gamma$, UKF-add-$\gamma$ and UKF-aug-$\gamma$ with reduced sigma point scaling represent competitive alternatives. The most convenient implementations might be the augmented unconstrained UKF-aug and UKF-fully-aug, which deliver acceptable estimations at low run times, even for nominal sigma point scaling. 

As a last note, the present study was limited to simulation data and a reduced AD model with linear outputs. In future studies, the presented UKFs must therefore be applied to real measurement data and higher-order AD models involving nonlinear output equations such as those proposed by \citep{Weinrich2021b} or \citep{Bernard2001}.

\section*{Acknowledgment}
The authors are thankful for funding from German Federal Ministry of Food and Agriculture of the junior research group on simulation, monitoring and control of anaerobic digestion plants (grant no. 2219NR333). S. H. thanks Maik Gentsch for his crucial advice and encouragement, and Julius Frontzek for the insights he contributed through cubature Kalman filter design.

\appendix
\newpage
\pdfbookmark[section]{Appendix}{sec:app} % sec:app is only the latex-intern label; Appendix is the name that the bookmark appears as; and [section] defines the level of the bookmark
\section*{Appendix: Model Equations}
%\section{Model Equations}
% 
In the following, the system equations of the model used in the present study are derived, cf. Section~\ref{sec:meth:ADModels}. Since it represents a vastly simplified form of the ADM1-R4 proposed by Weinrich and Nelles (2021) \citep{Weinrich2021b}, it is called ADM1-R4-Core. It describes the degradation of macro nutrients (carbohydrates, proteins and lipids) directly to biogas, i.e. methane (\chfour) and carbon dioxide (\cotwo), with the help of microbial biomass. 

Compared with the ADM1-R4, the following simplifications were made:
\begin{itemize}
	\setlength\itemsep{-5mm}
	\item Inorganic nitrogen $S_\mathrm{IN}$ was omitted because it acts as a quasi-autonomous state as described in \citep{Hellmann.2023b}. This means that its dynamics are only affected by other states, but inorganic nitrogen itself has no effect on the other states. Therefore, it can be deleted from the differential equation system without changing the dynamics of the remaining system. 
	\item The same holds for the ash concentration $X_\mathrm{ash}$ which was only added in \citep{Hellmann.2023b} to represent total and volatile solids measurements. 
	\item The gas phase was entirely omitted for two reasons: First, this results in linear output equations. Second, this allows to drop two additional states ($S_\mathrm{ch4,gas}$ and $S_\mathrm{co2,gas}$). 
	\item The mass concentrations of dissolved \chfour and \cotwo as well as microbial biomass were assumed to be directly measurable.
\end{itemize}
The state vector is comprised of the six mass concentrations
\begin{equation}
\label{eq:app:stateVector}
x= \left[S_\mathrm{ch4}, S_\mathrm{co2}, X_\mathrm{ch}, X_\mathrm{pr}, X_\mathrm{li}, X_\mathrm{bac}\right]^T. 
\end{equation}
The differential equations of ADM1-R4-Core read as follows 
\begin{subequations} \label{eq:app:R4-Core-ode}
\begin{align}
\dot x_1 &= c_1 \left(\xi_1 - x_1\right) u + a_{11} c_2 x_3 + a_{12} c_3 x_4 + a_{13} c_4 x_5 , \\
\dot x_2 &= c_1 \left(\xi_2 - x_2\right) u + a_{21} c_2 x_3 + a_{22} c_3 x_4 + a_{23} c_4 x_5 , \\
\dot x_3 &= c_1 \left(\xi_3 - x_3\right) u - c_2 x_3 + a_{34} c_5 x_6 , \\
\dot x_4 &= c_1 \left(\xi_4 - x_4\right) u - c_3 x_4 + a_{44} c_5 x_6 , \\ 
\dot x_5 &= c_1 \left(\xi_5 - x_5\right) u - c_4 x_5 + a_{54} c_5 x_6 , \\ 
\dot x_6 &= c_1 \left(\xi_6 - x_6\right) u + a_{61} c_2 x_3 + a_{62} c_3 x_4 + a_{63} c_4 x_5 - c_5 x_6. 
\end{align}
\end{subequations}
The measurement equations reduce to 
\begin{subequations}
	\label{eq:app:ADM1-R4-Core-mgl}
	\begin{align}
	y_1 &= S_\mathrm{ch4} = x_1 , \\
	y_2 &= S_\mathrm{co2} = x_2 , \\
	y_3 &= X_\mathrm{bac} = x_6,
	\end{align}
\end{subequations}
which can be put in vector-matrix form 
\begin{equation}
y = \begin{bmatrix}
y_1 \\ y_2 \\ y_3
\end{bmatrix} = 
\begin{bmatrix}
S_\mathrm{ch4} \\ S_\mathrm{co2} \\ X_\mathrm{bac}
\end{bmatrix} = 
\begin{bmatrix}
x_1 \\ x_2 \\ x_6
\end{bmatrix} = 
\begin{bmatrix}
1 & 0 & 0 & 0 & 0 & 0 \\
0 & 1 & 0 & 0 & 0 & 0 \\
0 & 0 & 0 & 0 & 0 & 1 
\end{bmatrix} \, x.
\end{equation}
Therein, $a_{ij}$ are the absolute values of the entries of the petersen matrix given in Tab.~\ref{tab:app:petersen-ADM1-R4-Core}, where $i$ denotes the column (component) and $j$ the row (process). For brevity, only those entries with an absolute value $\ne 1$ or $\ne 0$ were denoted with $a_{ij}$ specifically.
\begin{table}[h]
	\renewcommand{\arraystretch}{1.1}
	\setlength{\tabcolsep}{1pt}
	\caption{Petersen matrix of ADM1-R4-Core, derived from \citep{SorenWeinrich2017}.} \label{tab:app:petersen-ADM1-R4-Core} 
	\centering
	\begin{tabular}{ll*{6}{C{1.2cm}}l}
		\toprule \\ [-5mm]
		\multicolumn{2}{l}{\textbf{Component i $\rightarrow$}}  & 1  & 2 & 3 & 4 & 5 & 6 & \\ %[-4mm]
		\textbf{j} & \textbf{Process $\downarrow$}&   $S_\mathrm{ch4}$ & $S_\mathrm{co2}$ & $X_\mathrm{ch}$ & $X_\mathrm{pr}$ & $X_\mathrm{li}$ & $X_\mathrm{bac}$ & \textbf{Process rate} $r_j$\\ %[-0.3mm] 
		\midrule \\ [-5mm]
		1 & Fermentation $X_\mathrm{ch}$ & 0.2482 & 0.6809 & -1 & & & 0.1372 & $c_2 \,  x_3$\\ [1mm]
		2 & Fermentation $X_\mathrm{pr}$ & 0.3221 & 0.7954 & & -1 & & 0.1723 & $c_3 \,  x_4$\\ [1mm] 
		3 & Fermentation $X_\mathrm{li}$ & 0.6393 & 0.5817 & & & -1 & 0.2286 & $c_4 \,  x_5$\\ [1mm]
		\hdashline[0.6pt/1.8pt] \\ [-4mm]
		4 & Decay $X_\mathrm{bac}$ & & & 0.18 & 0.77 & 0.05 & -1 & $c_5 \,  x_6$\\ %[-1mm]
		\bottomrule		
	\end{tabular}
\end{table}

Table~\ref{tab:app:R4-Core-nomenclature} summarizes the model parameters $c_i$ of ADM1-R4-Core as well as the control variable. The dotted horizontal line should underline the different categories of parameters. $c_1$ is time invariant, whereas $c_2$ - $c_5$ are slowly time variant for varying substrates and subsequently adapting composition of the microbial community. During long-term process operation in practice, $c_2$ - $c_5$ need to be repeatedly updated throug parameter identification.
\begin{table}[h]
	\renewcommand{\arraystretch}{1.1}
	\centering
	\caption{Model parameters and control variable of ADM1-R4-Core according to Weinrich and Nelles (2021) \citep{Weinrich2021b} and in standard control notation. The stated true values apply for creation of synthetic measurement data and were taken from \citep{Weinrich.2021}. By contrast, the UKF values were used for Kalman filtering and were chosen differently to account for a plant model mismatch, see Section~\ref{sec:res:simulation-scenario}.}
	\label{tab:app:R4-Core-nomenclature}
	\begin{tabular}{clcc} 
		\toprule
		Control notation & Notation according to \citep{Weinrich2021b} & True value & UKF value\\
		\midrule
		$u$ & $q$ & \multicolumn{2}{c}{see Table~\ref{tab:meth:feedProperties}} \\ 
		% 
%		\hdashline[0.6pt/1.8pt] \\ [-4mm]
		\midrule
		$c_1$ & $V_\mathrm{liq}^{-1}$ & \multicolumn{2}{c}{\SI{0.01}{\per\liter}} \\ 
		\hdashline[0.6pt/1.8pt] \\ [-4.5mm] 
		$c_2$ & $k_\mathrm{ch}$ & \SI{0.25}{\per\hour} & \SI{0.3196}{\per\hour} \\
		$c_3$ & $k_\mathrm{pr}$ & \SI{0.20}{\per\hour} & \SI{0.2557}{\per\hour} \\
		$c_4$ & $k_\mathrm{li}$ & \SI{0.10}{\per\hour} & \SI{0.1278}{\per\hour} \\
		$c_5$ & $k_\mathrm{dec}$ & \SI{0.02}{\per\hour} & \SI{0.0256}{\per\hour} \\
		\bottomrule
	\end{tabular}
\end{table}

In \eqref{eq:app:R4-Core-ode}, $\xi_i$ denote the inlet concentrations of inflowing substrates. For the scope of this study, a mixture of cattle manure and maize silage was assumed as a substrate. The corresponding inlet concentrations were taken from \citep{Weinrich.2021} and are shown in Table~\ref{tab:app:R4-Core-xIn}. 
\begin{table}[htb]
	\centering
	\caption{Inlet concentrations $\xi_i$ for substrate mix of cattle manure and maize silage taken from \citep{Weinrich.2021}.}
	\label{tab:app:R4-Core-xIn}
	\begin{tabular}{c l S[table-format=1.6]	} 
		\toprule
		{\makecell[t]{Variable}} & {\makecell[t]{Corresponding component}} & {\makecell[t]{Value [\si{\kilogram\per\cubic\meter}]}} \\
		\midrule
		$\xi_1$ & $S_\mathrm{ch4}$ & 0 \\ [1mm]
		$\xi_2$ & $S_\mathrm{co2}$ & 0 \\ [1mm]
		$\xi_3$ & $X_\mathrm{ch}$ & 23.398 \\ [1mm]
		$\xi_4$ & $X_\mathrm{pr}$ & 4.750 \\ [1mm]
		$\xi_5$ & $X_\mathrm{li}$ & 1.381 \\ [1mm]
		$\xi_6$ & $X_\mathrm{bac}$ & 0 \\ [1mm]
		\bottomrule
	\end{tabular}
\end{table}

Lastly, measurement noise was considered by adding zero-mean Gaussian random numbers to nominal outputs obtained through \eqref{eq:app:ADM1-R4-Core-mgl}. Corresponding standard deviations were chosen as summarized in Table~\ref{tab:app:stdDevMeasNoise}. 
\begin{table}[H]
	\centering
	\caption{Standard deviations of additive measurement noise of individual measurements.}
	\label{tab:app:stdDevMeasNoise}
	\begin{tabular}{c l S[table-format=1.6]} 
		\toprule
		{\makecell[t]{Variable}} & {\makecell[t]{Corresponding output}} & {\makecell[t]{Value [\si{\kilogram\per\cubic\meter}]}} \\
		\midrule
		$\sigma_1$ & $y_1 = S_\mathrm{ch4}$ & 0.8 \\ [1mm]
		$\sigma_2$ & $y_2 = S_\mathrm{co2}$ & 1.0 \\ [1mm]
		$\sigma_3$ & $y_3 = X_\mathrm{bac}$ & 0.4 \\ [1mm]
		\bottomrule
	\end{tabular}
\end{table}

\interlinepenalty 10000	
% verhintert einen Seitenumbruch im Literaturverzeichnis innerhalb einer Quelle

\bibliographystyle{unsrt}	% Use this style if your thesis is in english (auch nutzbar, wenn et al. statt u.a. genutzt werden soll)
\clearpage
\bibliography{LiteraturUKF}							

\end{document}